\documentclass[aps,prx,superscriptaddress,notitlepage,twocolumn]{revtex4-2}
\usepackage{graphicx,enumerate,verbatim,bbold}
\usepackage{amsmath,amssymb,amsthm,float,mathrsfs}
\usepackage{dsfont}
\usepackage[colorlinks]{hyperref}
\usepackage[T1]{fontenc}    
\usepackage[utf8]{inputenc}
\usepackage{lmodern}
\usepackage{braket}
\usepackage{multirow}
\usepackage{bm}
\usepackage[caption=false]{subfig}
\usepackage{dcolumn}
\usepackage{hyperref}

\usepackage[normalem]{ulem}
\usepackage[usenames,dvipsnames]{color}
\newcommand{\bqa}{\begin{eqnarray}}
\newcommand{\eqa}{\end{eqnarray}}

\newcommand{\be}{\begin{equation}}
\newcommand{\ee}{\end{equation}}

\begin{document}

\title{Mitigating many-body quantum crosstalk with tensor-network robust control}
\author{Nguyen H. Le}
%\email[]{Your e-mail address}

\affiliation{School of Mathematics and Physics, University of Surrey, Guildford GU2 7XH, United Kingdom}
\affiliation{Blackett Laboratory, Imperial College London, SW7 2AZ, United Kingdom}
\affiliation{Qolumbus SAS, 78 Boulevard de la République, 92100 Boulogne-Billancourt, France}
\author{Florian Mintert}
\affiliation{Blackett Laboratory, Imperial College London, SW7 2AZ, United Kingdom}

\author{Eran Ginossar}
\affiliation{School of Mathematics and Physics, University of Surrey, Guildford GU2 7XH, United Kingdom}

%% ****** Abstract of paper ****** %
\begin{abstract}
Quantum crosstalk poses a major challenge to scaling up quantum computations as its strength is typically unknown and its effect accumulates exponentially as system size grows. Here, we show that many-body robust  control can be utilized to suppress unwanted couplings during multi-qubit  gate operations and state preparation. By combining tensor network simulations with the GRAPE algorithm, and leveraging an efficient random sampling over noise ensembles, our method overcomes the exponential scaling of the Hilbert space. We demonstrate its effectiveness for designing control solutions for high-fidelity implementations of parallel X gates and parallel CNOT on a  chain of 50 qubits, and for realizing a 30-qubit GHZ state and the ground state of a 20-qubit Heisenberg model. In the presence of many-body quantum crosstalk due to parasitic interaction between neighboring qubits, robust control results in order-of magnitude improvement in   fidelity for large system sizes. These findings pave the way for more reliable operations on near-term quantum processors.
\end{abstract} 
\maketitle

\section{Introduction}

In state‐of‐the‐art quantum computing architectures, ranging from superconducting circuits and trapped‐ion chains to semiconductor spin arrays and Rydberg atom arrays, quantum crosstalk, i.e., spurious “parasitic” interactions between qubits, presents a pervasive obstacle to achieving high‐precision control \cite{Zhou2023,Jirovec2025,Fang_crosstalk_2022,Kashyap2025,Madzik2025}. These residual couplings introduce spurious phase shifts, correlated errors, and unwanted entanglement which degrade gate fidelities and compromise error‐correction thresholds. As system sizes grow and control lines become more densely packed, the cumulative effect of weak but numerous parasitic terms can lead to substantial coherent leakage out of the computational subspace and accelerate decoherence \cite{Zhou2023,Zhao2022,niu_multi-qubit_2024}. Even when each coupling is individually small, their collective impact during multi‐qubit operations can overwhelm conventional calibration and compensation techniques, preventing the scalability and reliability of quantum processors. Addressing these parasitic interactions is therefore crucial for practical large‐scale quantum computation and  simulation.

Robust optimal control \cite{Weidner2025, koch2022quantum} offers a versatile framework for counteracting the unwanted effects of unknown perturbations in quantum hardware. By explicitly incorporating uncertain system parameters, one can design control pulses that maintain high fidelity across an ensemble of possible Hamiltonian realizations. Techniques like ensemble‐averaged GRAPE \cite{Khaneja2005} or Krotov’s method \cite{Reich2012} systematically sample over noise distributions and  minimize the worst‐case or average errors \cite{chen_robust_2025,Araki2023,Cykiert2024,Allen2017,Stefanescu2024}, yielding driving fields that are intrinsically insensitive to small deviations from the unpertubed model. Crucially, this approach does not require precise prior knowledge of every perturbation; rather, it exploits statistical information or bounded uncertainty ranges to build resilience into both state preparation and logic gates. As a result, robust optimal control can %significantly extend coherence times
improve fidelity \cite{Shapira_robust_2018, Webb_resilient_2018}, suppress coherent leakage \cite{Motzoi_simple_pulses_2009}, and relax hardware‐precision requirements \cite{Milne_phase_modulated_2020}, making it a promising strategy for pushing quantum devices toward fault‐tolerant performance.

However, the scaling of quantum devices poses substantial challenges for the field of optimal control.
The exponential growth of the Hilbert space renders techniques based on exact  state propagation impractical for large systems. While the general simulation of many-body dynamics is classically intractable, optimal solutions can still be obtained in special cases, for instance, when the optimal evolution remains confined to a low-entanglement sector of the Hilbert space \cite{Eisert2010}. In such scenarios, tensor-network (TN) methods provide an efficient representation of the dynamics and enable scalable optimization \cite{verstraete2008matrix, Cirac2021}.  Although TN-based optimal-control techniques employing both gradient-free \cite{doria2011optimal} and and gradient-based \cite{Jensen2021} strategies have been developed, particularly for controlling quantum phase transitions, robustness against variations in model parameters remains an open and largely unexplored problem.

Alternative non-TN approaches to robust many-body control have been proposed and applied to crosstalk mitigation, but they are typically limited to systems whose Hamiltonians generate small Lie algebras \cite{Stefanescu2024,liecontrol}. In this work, we introduce a non-perturbative robust optimal-control framework applicable to a broad class of many-body Hamiltonians. Our method combines matrix-product-state/operator simulations with the GRAPE algorithm and efficient random sampling over uncertain parameters. Because numerical optimization of many-body dynamics is computationally intensive, incorporating robustness, through ensemble averaging over parameter variations, can be especially demanding. We show, however, that averaging over a relatively small number of random realizations, scaling linearly with system size, is sufficient to achieve strong robustness. This substantially reduces the computational cost and enables the application of our approach to large systems. The full framework has been implemented as an open-source numerical library, described in the Appendix.

\begin{figure}[t]
    \centering
\includegraphics[width=0.9\linewidth]{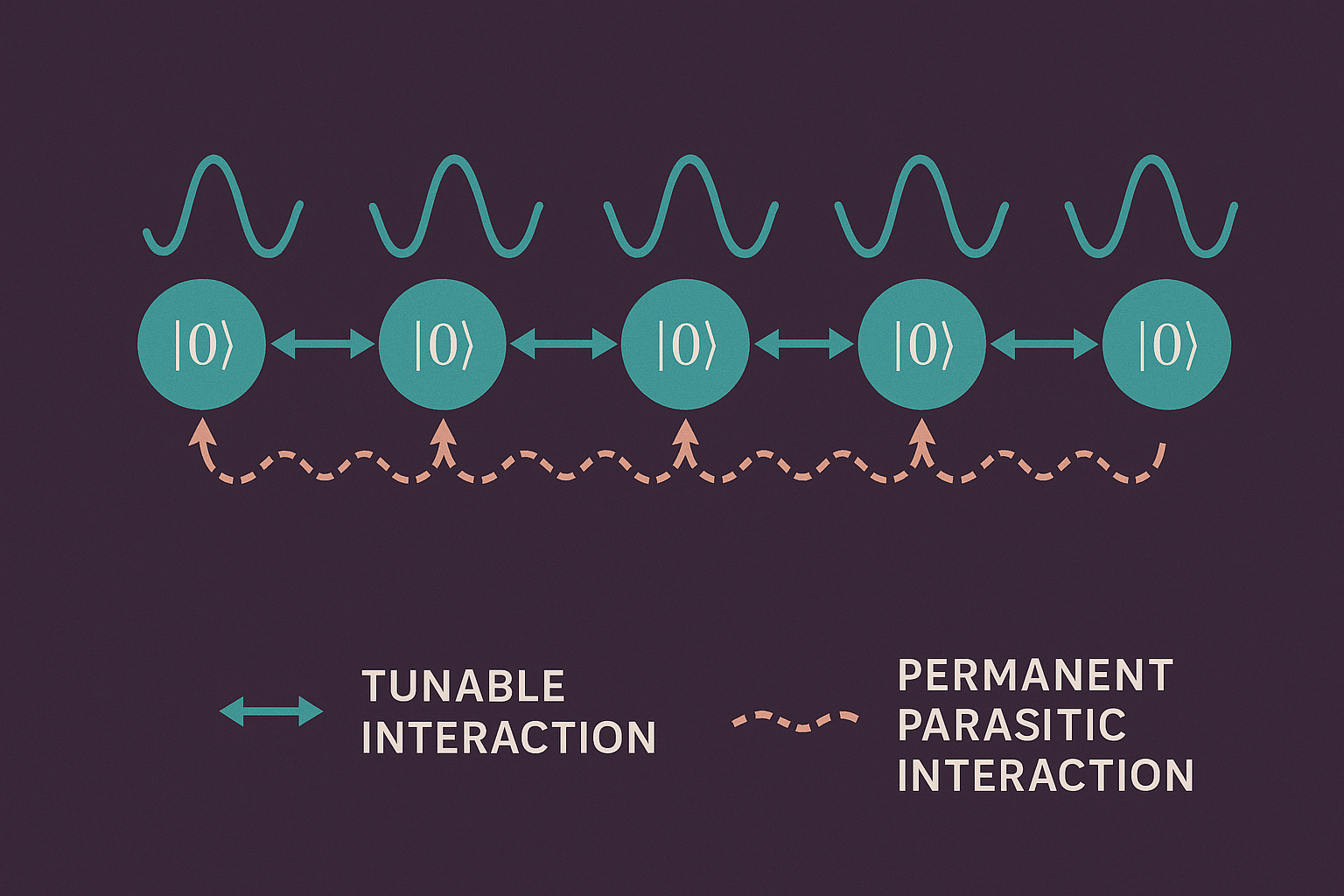}
\caption{Illustration of a qubit chain with tunable coupling and permanent parasitic interaction. Each qubit is controlled by a single-qubit drive.}
    \label{fig:qubitchain}
\end{figure}

The present TN-based method is applied to the design of high-fidelity many-body gates and state-preparation protocols in a one-dimensional (1D) chain of up to 50 qubits subject to the ubiquitous Heisenberg-type quantum crosstalk between neighboring qubits, described by  $J_xXX+J_yYY+J_zZZ$ interaction  (see Fig.~\ref{fig:qubitchain}). Despite unknown parasitic interaction strengths varying by up to 5\% of the system’s dominant energy scale, robust control reduces the infidelity of these many-body operations by up to two orders of magnitude.

The key steps of the robust control algorithm are presented in Sec. II. In Sec. III, we demonstrate its application to the robust implementation of parallel single-qubit and two-qubit gates in a linear qubit chain with many-body parasitic interactions. Section IV is devoted to the robust preparation of the GHZ state and the ground state of the Heisenberg model. Finally, Sec. V provides a discussion and outlook, including suitable experimental platforms and implications for near-term quantum information processing.

\section{Tensor-network robust  control}

In optimal control of quantum dynamics, the unitary time-evolution operator  $U(t)$ obeys the Schr\"odinger equation
\begin{align}\label{eq:sde}
    \frac{\partial}{\partial t} U(t)=-i H(t)U(t), 
\end{align}
where $H(t)$ is the time-dependent system Hamiltonian. The control problem consists of iteratively refining the time dependence of $H(t)$ such that the final evolution operator $U(T)$ closely approximates a desired multi-qubit quantum gate described by a target unitary $U_T$.  

For many-body systems, an exact representation of $U(t)$ in the computational basis requires an exponentially large number of coefficients. To overcome this limitation, we employ a matrix-product-operator (MPO) representation. Instead of explicitly storing all expansion coefficients $c_{i_1j_1,i_2j_2,\dots,i_nj_n}$ of an $n$-body operator
\be
U=\sum_{\{i_k,j_k\}} c_{i_1j_1,i_2j_2,\dots,i_nj_n}\ket{i_1}\!\bra{j_1}\otimes \ket{i_2}\!\bra{j_2}\otimes \dots \otimes \ket{i_n}\!\bra{j_n},
\ee
the MPO formalism factorizes these coefficients as
\be
c_{i_1j_1,i_2j_2,\dots,i_n j_n }=A^{[1]}_{i_1j_1}A^{[2]}_{i_2j_2}\dots A^{[n]}_{i_n j_n},
\ee
where $A^{[k]}_{i_k j_k}$ are elements of the matrices $A^{[k]}$ , with the first and last  being row and column vectors, respectively. Any $n-$body operator can be represented exactly in this form provided that the bond size $D$, {\it.i.e.}, the size of the matrices $A_{i_k,j_k}^{[k]}$, is sufficiently large. Crucially, however, many physically relevant operators can be accurately approximated with relatively small bond sizes, reducing the computational cost from exponential to sub-exponential scaling.

Starting from the initial operator $U(0)\equiv\mathbb{1}^{\otimes n}$, represented as a matrix product operator (MPO), the time evolution is simulated using the time-evolving block decimation (TEBD) method \cite{vidal_efficient_2004,paeckel_time-evolution_2019}. In TEBD, the propagator over a small time step is approximated by a Trotter decomposition into single- and two-qubit gates, which can be efficiently contracted into the MPO representation. For a 1D system with nearest neighbor interactions $V_{j,j+1}$  and local single-qubit  controls $X_j$ and $Y_j$, the Trotterized evolution operator takes the form $U(T)=\prod_l U_l$ where each time-slice $U_l$ is given by
\begin{equation}\label{eq:circuit}
\prod_{j=1}^n e^{-i \Delta t x_{jl} X_j} \prod_{j=1}^n e^{-i \Delta t y_{jl} Y_j}\prod_{j=1}^{n-1}e^{-ig_j\Delta t V_{j,j+1}}.
\end{equation}
Here, $\Delta t$ is a small time step, and $x_{jl}$, $y_{jl}$ denote the control amplitudes applied to qubit $j$ at time $t_l$, and $g_j$ the interaction strengths. Contracting each $U_l$ to the MPO generally increases its bond size. To control computational complexity, the MPO is compressed after each time step to a predetermined maximum bond size $D_{\text{max}}$ using truncated singular-value decomposition  (SVD) \cite{Hubig2017}. An important advantage of the Trotterized form in Eq.~\eqref{eq:circuit} is that its exact gradients with respect to the control amplitudes $x_{jl}$ and $y_{jl}$ are known analytically \cite{Jensen2021} and can be computed efficiently via MPO contractions.

The figure of merit of the optimization is the gate infidelity,
\be\label{eq:gateinfid}
\chi=1-\vert \mbox{tr}(U(T)^{\dagger}U_T)/2^n\vert^2,
\ee
which quantifies the deviation between the final unitary $U(T)$ and the target unitary $U_T$. In the presence of parasitic interaction terms whose strength is unknown or subjected to static noise, the system Hamiltonian $H(t)$ and the resulting dynamics become distributed over an ensemble of realizations.We model this uncertainty by assuming that the strengths $J_j$ of the parasitic interactions are independent random variable uniformly distributed in the interval $[-\Delta J, \Delta J]$, where $\Delta J$ denotes the maximum error magnitude. Consequently, the final unitary $U(T)$ is itself ensemble-distributed. For achieving robustness against variation within this ensemble, the ensemble-averaged infidelity,
\be
\bar{\chi}=1- \frac{1}{M}\sum_{s=1}^M |\mbox{tr}(U^{(s)}(T)^{\dagger}U_T)/2^n\vert^2,
\ee
is minimized, where $U^{(s)}(T)$ denotes the final unitary corresponding to the $s-$th sampled realization and $M$ is the total number of sampling points in the ensemble.

The state-preparation problem can be treated in a similar fashion. The quantum state is first represented as a matrix-product state (MPS) and propagated in time using the TEBD algorithm. Robust control is then achieved by minimizing the ensemble-averaged state infidelity,
\be
\bar{{\cal I}}=1- \frac{1}{M}\sum_{s=1}^M \vert \langle \psi^{(s)}(T)\vert \psi_T\rangle \vert^2,
\ee
which measures the average deviation between the final state $\ket{\psi^{(s)}(T)}$ obtained for the $s-$th ensemble realization and the target state $\ket{\psi_T}$.

Although MPS and MPO representations enable efficient simulation of many-body dynamics, optimal control remains computationally demanding due to the large number of required optimization iterations and the cost of evaluating gradients. Incorporating robustness further increases this complexity, as it necessitates optimizing an ensemble-averaged objective function. In many-body systems, the number of unknown parasitic interaction strengths typically grows polynomially with system size, but the volume of the parameter space from which the Hamiltonian is drawn grows exponentially with the number of unknown parameters. Empirical results in this work indicate that it is sufficient to optimize using an ensemble whose size scales only linearly with the number of unknown parameters. Specifically, a two-stage strategy is employed: optimization is first carried out over an ensemble of size $M = 2n_J$, where $n_J$ denotes the number of parasitic interaction terms; the mean infidelity is then estimated using a larger ensemble of $5M$ samples, which is verified to be sufficiently large for convergence of the ensemble average.

The infidelity for each ensemble realization is evaluated in parallel across distributed computing nodes. Analytic expressions for the exact gradients are derived (see Appendix) and efficiently computed via tensor contractions. The resulting ensemble-averaged infidelity is minimized using an L-BFGS quasi-Newton optimization algorithm.

Owing to the high-dimensional control landscape of many-body systems, particularly for tasks involving $\sim 10^4$ control parameters considered in this paper, shallow local minima are ubiquitous and are frequently encountered when optimization is initialized with random control fields. We find that sequential optimization over system parameters is highly effective in mitigating this issue. Specifically, the optimized control solution for a system of size $n-1$ is used as the initial guess for optimizing the system of size 
$n$. Similarly, robustness is introduced incrementally: the solution optimized in the absence of errors is used as the initial guess for the 1\% error case, which then seeds the 2\% error optimization. This procedure is repeated in 1\% increments until the desired error tolerance is reached.

\section{Robust gates}
Consider a chain of identical qubits with tunable nearest-neighbor $ZZ$ coupling and local single qubit drives along the $X$ and $Y$ quadratures. In the rotating-wave approximation, the system Hamiltonian is 
\be\label{eq:sysH}
H(t)=\sum_{j=1}^n \left(x_j(t) X_j +y_j(t) Y_j \right) +\sum_{j=1}^{n-1}g_j(t) Z_jZ_{j+1}+H_p,
\ee
where $x_j(t)$ and $y_j(t)$ denote the envelope functions of the single-qubit drives, $g_j$ is the tunable two-qubit interaction strength, and $H_p$ represents parasitic interactions responsible for quantum crosstalk. The focus of this work 
is on the  general Heisenberg-type parasitic interaction,
\be\label{eq:Hp}
H_p=\sum_{j=1}^{n-1} \left(J^x_j X_jX_{j+1} + J^y_j Y_jY_{j+1} + J^z_j Z_j Z_{j+1}\right),
\ee
which captures the dominant forms of parasitic couplings encountered in a wide range of quantum hardware platforms.

The objective is to realize single-qubit and two-qubit gates that are robust against such quantum crosstalk. In the ideal system without $H_p$, a global $\pi$ pulse along the X quadrature implements the parallel $X$ gate,
\be
U_X=\prod_{j=1}^n X_j.
\ee
Similarly, by activating interactions only between alternating pairs of neighboring qubits, 
\be
g_j=g \ \text{for  $j$ odd},
\quad g_j=0 \ \text{for $j$ even},
\ee
one can straightforwardly design control pulses that realize parallel CNOT gates,
\be
U_{\text{C}}=\prod_{j=1}^{n/2} \text{CNOT}_{2j-1,2j}.
\ee
\begin{figure}[t]
    \centering    \includegraphics[width=0.9\linewidth]{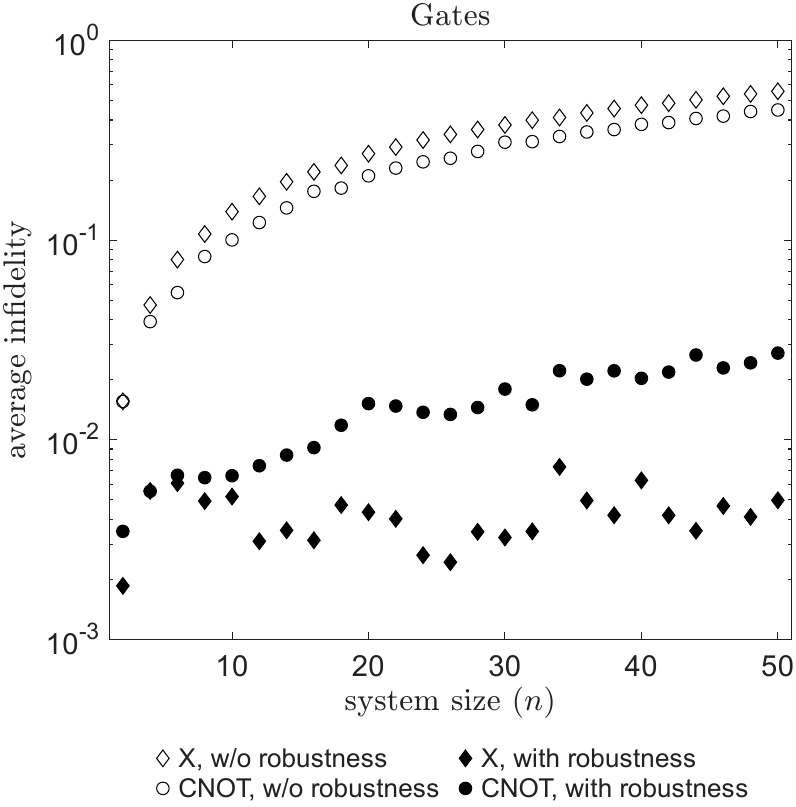}
\caption{Ensemble-averaged infidelity for the parallel 
$X$ gate (diamonds) and parallel CNOT gate (circles). Results are shown for control solutions optimized without robustness (open symbols) and with robustness against quantum crosstalk at 
5\% of the main energy scale (solid symbols). Without robustness, the average infidelity increases rapidly with system size, approaching 0.5 at $n=50$. In contrast, robust optimization suppresses the infidelity by nearly three orders of magnitude for the parallel $X$ gate and by approximately two orders of magnitude for the parallel CNOT. The maximum MPO bond size during optimization is $D_{\text{max}}=20$, and the pulse duration is independent of system size.}
    \label{fig:gate}
\end{figure}

The target unitaries $U_X$ and $U_{\text{C}}$ admit compact MPO representations with bond sizes 2 and 4, respectively. In the absence of parasitic interactions, these correspond to independent single- and two-qubit control problems. However, the presence of $H_p$
 couples all neighboring qubits, transforming the task into a genuine many-body control problem. During the evolution, the MPO representing the system dynamics generally acquires a bond size larger than that of the initial and target operators.

The optimization target is the gate infidelity defined in Eq.~\eqref{eq:gateinfid}, averaged over an ensemble of parasitic Hamiltonian realizations. Figure~\eqref{fig:gate} shows the ensemble-averaged infidelity for parasitic interaction strengths equal to 5\% of the main energy scale, which is the Rabi frequency $\Omega_{\pi}$ of the  $\pi$ pulse for the parallel $X$ gate and the interaction strength $g$ for the parallel CNOT.
\begin{figure*}[t]
    \centering   \includegraphics[width=0.8\linewidth]{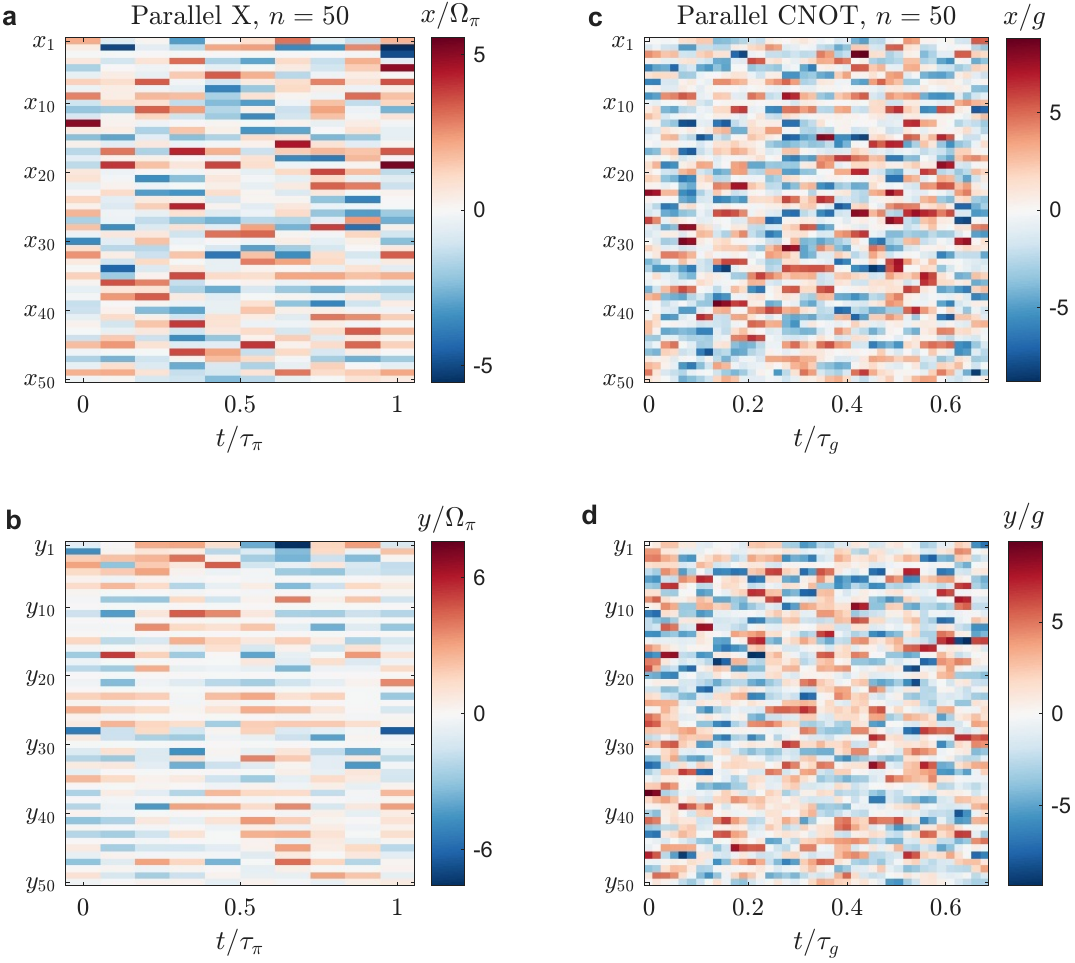}
\caption{The pulses optimized with robustness for a chain of $n=50$ qubits. Panels (a,b) show the pulses for the parallel $X$ gate, and panels (c,d) show those for the parallel CNOT.  Time is given in units of $\tau_{\pi}$ and $\tau_g=2\pi/g$, and amplitudes are shown in units of $\Omega_{\pi}$ and $g$, respectively. For each case, all 50 control pulses on the X-quadrature (top) and Y-quadrature (bottom) are displayed, with each pulse shown as a row of color-coded amplitudes. The resulting many-body gate infidelities are approximately $5.0\times 10^{-3}$ for the parallel $X$ gate and $2.7\times 10^{-2}$ for the parallel CNOT, despite parasitic interaction at 5\% of the main energy scales.}
    \label{fig:pulse_gate}
\end{figure*}

 Control solutions optimized without robustness, i.e, optimized with $H_p=0$, exhibit rapidly increasing infidelity with system size in the presence of the parasitic $H_p$, reaching approximately 50\% at $n=50$.
 In contrast, robust optimization suppresses the many-body infidelity to approximately $5.0\times 10^{-3}$ for the parallel $X$ gate and $2.7\times 10^{-2}$ for the parallel CNOT. These values correspond to the infidelity of a large number of gates executed in parallel. The inferred infidelity per gate, $1-(1-\bar{\chi})^{1/n}$, where $\bar{\chi}$ is the average many-body infidelity, for the parallel $X$ gate and $1-(1-\bar{\chi})^{2/n}$ for the parallel CNOT, is significantly smaller, yielding values of approximately $1.0\times 10^{-4}$
 and $1.1\times 10^{-3}$, respectively.

Robust optimization is performed by minimizing the ensemble-averaged infidelity over $ M=6(n-1)$ uniformly sampled realizations of the parasitic interaction. The resulting control pulses are then validated using a larger ensemble containing $5M$ samples. Throughout the evolution, the MPO bond size is restricted to $D_{\text{max}}=20$. The duration of the optimized pulses is independent of system size: $T\sim\tau_{\pi}$ for the parallel $X$ gate and $T\sim\tau_g/2$ for the parallel CNOT, where $\tau_{\pi}$ is the duration of the $\pi$ pulse and $\tau_g=2\pi/g$ is the characteristic interaction timescale. This system-size-independent duration reflects the parallel nature of the system dynamics owing to the absence of long-range correlations in the target unitaries.

The optimized control pulses for $n=50$ qubits are shown in Fig.~\ref{fig:pulse_gate}. All 100 single-qubit control functions, $x_j(t)$ and $y_j(t)$ with $1 \le j \le 50$, are displayed, each as a row of color-coded amplitudes. These are piece-wise pulses with 10 and 20 time bins for the parallel $X$ and CNOT gate, respectively. For the parallel $X$ gate, the pulse duration remains comparable to that of a single $\pi$ pulse, demonstrating that the target unitary can be reached in parallel despite the presence of entangling parasitic interactions.  The maximum amplitudes, however, are approximately 5 times the Rabi frequency of the $\pi$ pulse for the X-quadrature, and 6 times for the Y-quadrature. This significant increase in amplitude is necessary for keeping the unitary dynamics robust against variations in the parasitic interaction strength $J^{x,y,z}_j$.

For the parallel CNOT, the pulse duration is on the order of $\tau_g$, and the maximum control amplitudes are approximately eight times the interaction strength $g$. As in the single-qubit case, these enhanced amplitudes are essential for maintaining robustness against variations in the parasitic couplings. Such amplitude requirements are well within current experimental capabilities, as single-qubit Rabi frequencies exceeding an order of magnitude over two-qubit interaction strengths are routinely achieved in superconducting qubits~\cite{kim_evidence_2023}, trapped ions~\cite{Ballance2016}, spin qubits~\cite{edlbauer_11-qubit_2025}, and neutral-atom platforms~\cite{bluvstein_logical_2024}.

\section{State preparation}
As a second application, the effectiveness of the present robust control algorithm is demonstrated for the robust preparation of many-body quantum states relevant to quantum information processing and many-body physics. The 1st example is the preparation of the Greenberger–Horne–Zeilinger (GHZ)  state in a 1D chain of qubits with  nearest-neighbor ZZ interactions and single-qubit control, described by the Hamiltonian in Eq.~\eqref{eq:sysH} with 
\be
g_j=g \quad \forall j.
\ee
As before, quantum crosstalk arising from the parasitic interaction in Eq.~\eqref{eq:Hp} is included.

The system is initialized in the product state $\ket{\psi(0)}=\ket{0}^{\otimes n}$ which admits a trivial MPS representation with bond size 1, given by $A_{0}^{[j]}=1$ and $A_{1}^{[j]}=0$ for $1\leq j\leq n$. The target GHZ state, $\ket{\psi}_{\text{GHZ}}=(\ket{0}^{\otimes n}+\ket{1}^{\otimes n})/\sqrt{2}$, has an exact MPS representation with bond size 2:
\begin{align}
&A^{[1]}_0=\begin{pmatrix} 2^{-1/2} \
0 \end{pmatrix}, \quad A_1^{[1]}=\begin{pmatrix} 0  \ 2^{-1/2} \end{pmatrix}, \\ \nonumber
&A^{[n]}_0=\begin{pmatrix} 1 \\ 
0 \end{pmatrix}, \quad A_1^{[n]}=\begin{pmatrix} 0 \\
1 \end{pmatrix} ,
\end{align}
and for the intermediate sites $\ 2\leq j\leq n-1$
\begin{align}
    &A_0^{[j]}= \begin{pmatrix} 1 &0 \\
0  &0\end{pmatrix}, \quad
A_1^{[j]}=\begin{pmatrix} 0 &0\\
0 &1\end{pmatrix}.
\end{align}

The present method is not restricted to target states with simple analytical MPS representations. As a second example, the ground state of the anti-ferromagnetic Heisenberg model, 
\be
H_{T}=\sum_{j=1}^{n-1}(X_jX_{j+1}+Y_jY_{j+1}+Z_jZ_{j+1}),
\ee
is chosen as the target state. This model is gapless in the thermodynamic limit. For even system sizes, however, the model exhibits a finite-size gap that decreases rapidly with increasing $n$, resulting in a unique but increasingly complex ground state. This state does not admit a simple analytical MPS form and is therefore obtained numerically using variational MPS techniques~\cite{verstraete2008matrix}. A bond size of $D=20$ is verified to be sufficient for an accurate representation of the ground state for system sizes up to $n=20$.
\begin{figure}[t]
    \centering
\includegraphics[width=0.9\linewidth]{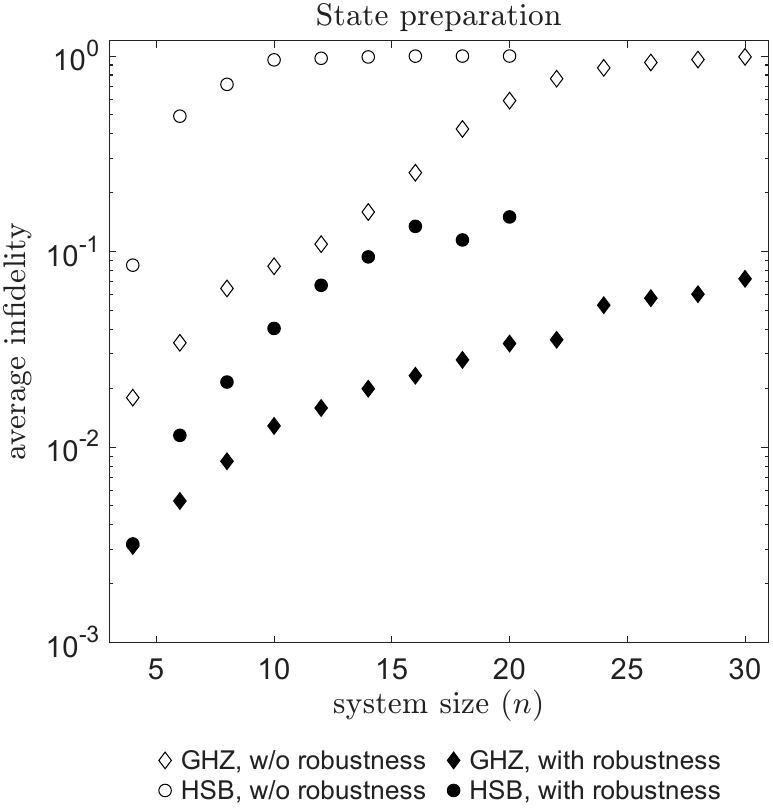}
\caption{Ensemble-averaged infidelity for the preparation of the GHZ state (diamonds) and the Heisenberg ground state (circles). Results are shown for control solutions optimized without robustness (open symbols) and with robustness against parasitic interactions at $5\%$ of the interaction strength $g$ (solid symbols). Without robustness, the infidelity increases exponentially with system size before saturating near unity. Robust optimization keeps the infidelity below $0.1$ for the GHZ state at $n=30$ and below $0.15$ for the Heisenberg ground state at $n=20$. The maximum bond size during optimization is $D_{\text{max}}=10$ for the GHZ state and $D_{\text{max}}=20$ for the Heisenberg ground state.}
    \label{fig:state}
\end{figure}
\begin{figure*}[t]
    \centering
\includegraphics[width=0.82\linewidth]{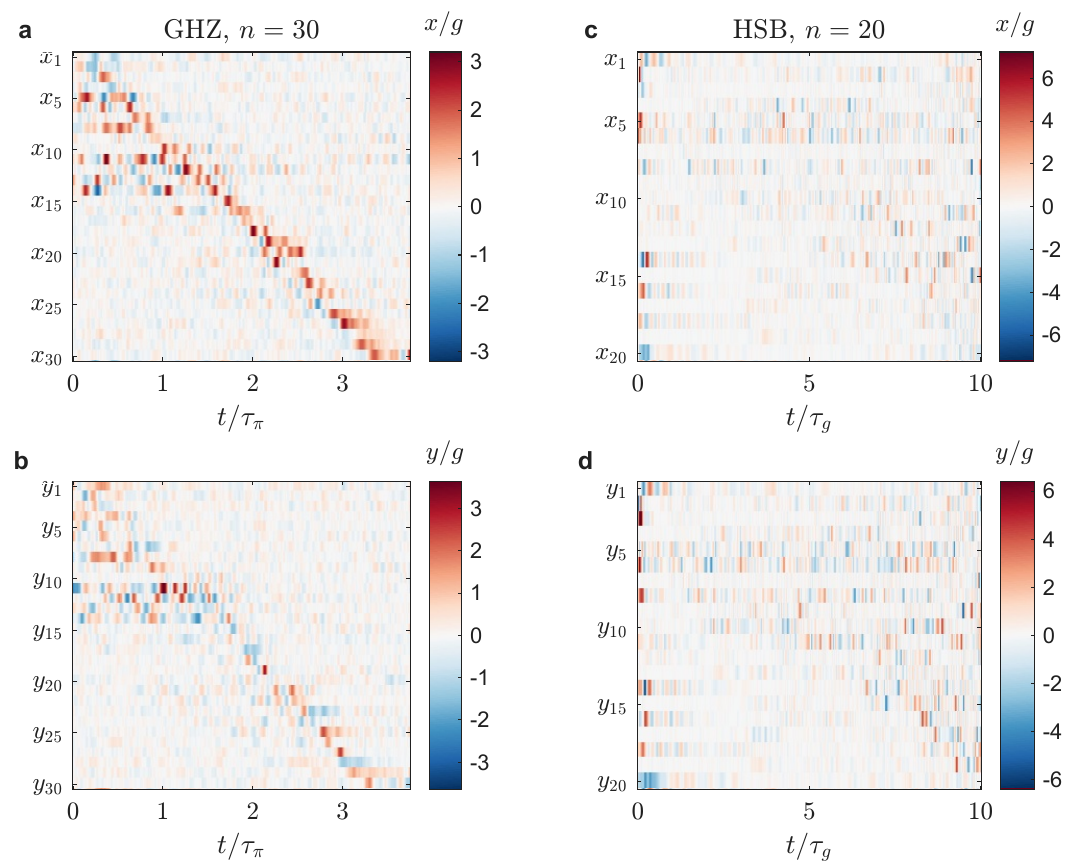}
\caption{
The pulses optimized with robustness for the preparation of the GHZ state with $n=30$ qubits (a,b) and the Heisenberg ground state with $n=20$ qubits (c,d). Pulse amplitudes, shown in units of $g$, are color-coded as functions of time in units of the interaction timescale $\tau_g = 2\pi/g$. Pulses on the $X$-quadrature ($Y$-quadrature) are shown in the top (bottom) panels, with each row corresponding to a single qubit. The resulting state infidelities are approximately $0.1$ for the GHZ state and $0.15$ for the Heisenberg ground state despite parasitic interactions at $5\%$ of the main interaction strength.}
    \label{fig:pulse_state}
\end{figure*}

For many applications involving the observation of quantum phase signatures such as long-range correlations, ultra-high fidelity is not required. Fidelities on the order of $ \gtrsim 50\% $ are often sufficient~\cite{omran2019generation,song2019generation,bernien_probing_2017}. While the preparation of entangled states with long-range correlations is more challenging than the parallel-gate examples discussed in Sec.~III, the fidelity requirements are less stringent.

Figure~\ref{fig:state} shows the ensemble-averaged infidelity for control solutions optimized with and without robustness against parasitic interactions at $5\%$ of the interaction strength $g$. Even at the smallest system size $n=4$, the effect of static noise is significant for both target states. As the system size increases, noise accumulates, leading to an exponential growth of the infidelity before saturation near unity. In contrast, robust optimization maintains substantially lower infidelities, reaching approximately $0.1$ for the GHZ state at $n=30$ and $0.15$ for the Heisenberg ground state at $n=20$. For large system sizes, the infidelity achieved with robust control is noticeably higher for the Heisenberg ground state than for the GHZ state. This difference can be attributed to the rapidly closing finite-size energy gap of the Heisenberg model, resulting in increasing entanglement in the ground state. For example, at $n=20$, this gap is only approximately 
2.5\% of the ground-state energy.

As in the robust-gate examples, the ensemble-averaged infidelity is minimized over $M = 6(n-1)$ uniformly distributed random samples of the parasitic interaction, and the final infidelity is evaluated using a larger ensemble of size $5M$. The maximum bond size is set to $D_{\text{max}}=10$ for the GHZ state and $D_{\text{max}}=20$ for the Heisenberg ground state. Due to the presence of long-range correlations in both target states, achieving low infidelity requires control pulses whose duration increases with system size. In this work, the pulse duration is chosen as $T \approx n\tau_g/8$ for the GHZ state and $T \approx n\tau_g/2$ for the Heisenberg ground state, where $\tau_g = 2\pi/g$ is the interaction timescale. These scalings are determined empirically.

The main factors limiting the achievable infidelity are the number of optimization iterations, fixed at 1000, and the number of time bins used in the piecewise-constant control pulses. For both target states, the number of time bins is increased linearly with system size, with $20n$ bins used in the simulations. Under these conditions, increasing the bond size beyond the chosen values does not lead to a noticeable improvement in fidelity, indicating that further gains would require longer pulse durations, finer temporal resolution, or additional optimization iterations.

The optimized pulses for the robust preparation of the 30-qubit GHZ state are shown in Figs.~\ref{fig:pulse_state}a and~\ref{fig:pulse_state}b. The prominent diagonal structure indicates a sequential driving pattern, where qubits further to the right in the chain are addressed later in time. This behavior is consistent with optimal quantum circuits for GHZ-state preparation using nearest-neighbor interactions~\cite{cruz_efficient_2019}, %\flo{can we give a citation?}\nguyen{citation added}
suggesting that the optimization algorithm is capable of discovering near-optimal circuit structures. Despite the presence of always-on many-body interactions, the required control amplitudes remain moderate, with maximum values of approximately $3g$.

Figures~\ref{fig:pulse_state}c and~\ref{fig:pulse_state}d show the optimized pulses for preparing the 20-qubit Heisenberg ground state. The upper triangle shape of the pulse structure again suggests a sequential protocol, though less pronounced than in the GHZ case. The maximum control amplitudes are similarly moderate, reaching approximately four times the interaction strength $g$.  

\section{Discussion and outlook}

The robust many-body optimal-control framework developed in this work is directly applicable to a range of current quantum simulation and quantum information experiments in which quantum crosstalk due to residual interactions constitutes a dominant source of error. In particular, the method is well suited for systems where interactions are short-ranged but imperfectly calibrated or controlled, and where permanent parasitic couplings accumulate as system size increases.

A particularly promising application of the present framework is in arrays of superconducting qubits, where quantum crosstalk arising from parasitic $ZZ$ interactions is a  persistent limitation. In transmon- and fluxonium-based architectures, residual static 
$ZZ$ couplings between neighboring qubits generate correlated phase errors that reduce gate fidelities  \cite{alghadeer_low_2026,Ni2022ScalableZZ,Mundada2019SuppressionZZ}. These interactions are difficult to fully remove through calibration and  become more severe as system size increases. By explicitly incorporating uncertainty in interaction strengths, the robust control approach developed here offers a systematic way to suppress such coherent errors without requiring precise knowledge of all parasitic couplings. The system sizes explored in this work are directly relevant to current superconducting processors, suggesting that tensor-network-based robust control could significantly enhance gate and state-preparation performance in these devices.

The framework is similarly applicable to other leading quantum computation and quantum simulation platforms.   In Rydberg atom arrays, residual van der Waals interactions and blockade imperfections introduce quantum crosstalk \cite{browaeys_many-body_2020,Simard2025,Zhang2012,Su2023}. In trapped ions, the spillover of focused laser beams on spectator qubits leads to parasitic interaction, a major error source limiting the fidelity of entangling gates \cite{Fang_crosstalk_2022,Kashyap2025}. Likewise, semiconductor quantum-dot arrays and spin-based quantum processors suffer from residual exchange coupling between neighboring spins, leading to unwanted entanglement and correlated errors \cite{Heinz2024,duan_mitigating_2025,Jirovec2025}. In all these systems, parasitic interactions are often only partially characterized and can drift over time, making exhaustive calibration  impractical. As demonstrated in this work, robust pulse shaping can substantially mitigate the resulting coherent errors even under such limited prior knowledge. 

The proposed method is also well suited to emerging modular quantum processor architectures. In these systems, individual modules typically contain  
10-100 qubits that can be locally controlled with high precision, while inter-qubit or inter-module couplings are mediated through slower and noisier channels \cite{edlbauer_11-qubit_2025,Li2024}. The problem sizes addressed here fall within this regime, providing an effective toolbox for high-fidelity gate implementation and state preparation within individual modules. By reducing sensitivity to coherent crosstalk, this approach can relax hardware-precision requirements and simplify calibration at the module level, supporting scalable quantum architectures.

\subsection*{Acknowledgements}
This work was supported by the U.K. Engineer
ing and Physical Sciences Research Council via the
EPSRC Hub in Quantum Computing and Simulation
(EP/T001062/1), the UK Innovate UK (project number 10004857), and the EPSRC strategic equipment grant no. EP/L02263X/1. The authors thank Modesto Orozco-Ruiz for suggesting the visualization style used in Figs.~3 and 5. 
\appendix
\section{Gradients}
The fidelity between the target state $\ket{\psi_T}$ and the final state $\ket{\psi(T)}$ is the absolute square of the overlap
\begin{equation}
O =\langle\psi_T\vert\psi(T)\rangle  = \langle b_{l+1} | U_l | a_{l-1} \rangle ,
\end{equation}
where $|a_{l-1}\rangle$ denotes the initial state propagated forward in time up to slice $l-1$,  
$|b_{l+1}\rangle$ is the target state propagated backward in time to slice $l+1$~\cite{le_scalable_2023},  
and $U_l$ is the time-evolution operator acting during time slice $l$, as defined in Eq.~\eqref{eq:circuit}.

The exact gradients of this overlap with respect to the control amplitudes applied at time slice $l$ are given by
\begin{align}
\frac{\partial O}{\partial x_{jl}}
&= \langle b_{l+1} | X_j U_l | a_{l-1} \rangle
= \langle b_{l+1} | X_j | a_l \rangle , \nonumber \\
\frac{\partial O}{\partial y_{jl}}
&= \langle b_{l+1} | Y_j U_l | a_{l-1} \rangle
= \langle b_{l+1} | Y_j | a_l \rangle ,
\end{align}
where $|a_l\rangle = U_l |a_{l-1}\rangle$ is the forward-propagated state at the end of time slice $l$.
These expressions allow the exact gradients to be evaluated efficiently using tensor-network contractions.

The gradients of the operator infidelity can be computed similarly by replacing the propagated state vectors with propagated operators and replacing the state overlap with the corresponding trace distance.

\section{A library for tensor-network optimal quantum control}
ToQC is a numerical library developed in \textsc{Matlab} to implement the tensor-network-based robust optimal control framework introduced in this work. The library and documentations are openly available on GitHub~\cite{toqc}.

The \texttt{mpo-unitary} branch provides routines for optimal control of unitary operators using MPO representations, while the \texttt{mps} branch supports optimal control of quantum states represented as MPS. Parallelized functions for computing ensemble-averaged infidelities and their gradients are identified by the suffix \texttt{prl}.
\\
\subsection*{Data Availability}
The simulation data
for this work are openly available \cite{zenodo}
\bibliography{refs.bib}
\end{document}